\documentclass[aps,prb,twocolumn,superscriptaddress,longbibliography]{revtex4-2}

\usepackage{amsmath,amssymb,bm,graphicx}
\usepackage{dcolumn}
\usepackage{hyperref}
\hypersetup{colorlinks=true,citecolor=blue,linkcolor=blue,urlcolor=blue}
\usepackage{braket}
\usepackage{physics}
\usepackage{xcolor}
\usepackage[utf8]{inputenc}
\usepackage[T1]{fontenc}
\usepackage{siunitx}
\usepackage{comment}
\usepackage{nicefrac}
\usepackage[normalem]{ulem}
\usepackage[english]{babel}
\newcommand{\AK}[1]{\textcolor{black}{#1}}

\newcommand{\AN}[1]{\textcolor{black}{#1}}

\newcommand{\MM}[1]{\textcolor{black}{#1}}

\newcommand{\DN}[1]{\textcolor{black}{#1}}

\begin{document}

\title{Kardar-Parisi-Zhang universality in optically induced lattices of exciton-polariton condensates}

\author{D.~Novokreschenov}
\affiliation{Abrikosov Center for Theoretical Physics, MIPT, Dolgoprudnyi, Moscow Region 141701, Russia}
\affiliation{Russian Quantum Center, Skolkovo, Moscow, 121205, Russia}

\author{V.~Neplokh}
\affiliation{Abrikosov Center for Theoretical Physics, MIPT, Dolgoprudnyi, Moscow Region 141701, Russia}
\affiliation{Russian Quantum Center, Skolkovo, Moscow, 121205, Russia}

\author{M.~Misko}
\affiliation{Hybrid Photonics Laboratory, Skolkovo Institute of Science and Technology, 121205 Moscow, Russia}

\author{N.~Starkova}
\affiliation{Hybrid Photonics Laboratory, Skolkovo Institute of Science and Technology, 121205 Moscow, Russia}

\author{T.~Cookson}
\affiliation{Hybrid Photonics Laboratory, Skolkovo Institute of Science and Technology, 121205 Moscow, Russia}

\author{A.~Kudlis}
\email{andrewkudlis@gmail.com}
\affiliation{Science Institute, University of Iceland, Dunhagi 3, IS-107 Reykjavik, Iceland}

\author{A.~Nalitov}
\affiliation{Abrikosov Center for Theoretical Physics, MIPT, Dolgoprudnyi, Moscow Region 141701, Russia}

\author{I. A. Shelykh}
\affiliation{Science Institute, University of Iceland, Dunhagi 3, IS-107 Reykjavik, Iceland}

\author{A. V.~Kavokin}
\affiliation{Abrikosov Center for Theoretical Physics, MIPT, Dolgoprudnyi, Moscow Region 141701, Russia}
\affiliation{Russian Quantum Center, Skolkovo, Moscow, 121205, Russia}
\author{P.~Lagoudakis}
\affiliation{Hybrid Photonics Laboratory, Skolkovo Institute of Science and Technology, 121205 Moscow, Russia}

\date{\today}

\begin{abstract}
We investigate space-time coherence in one-dimensional lattices of exciton-polariton condensates formed by fully reconfigurable non-resonant optical pumping. Starting from an driven-dissipative Gross-Pitaevskii equation with deterministic reservoir kinetics and stochastic condensate noise, we derive a discrete complex-field model that incorporates coherent tunnelling, reservoir-mediated dissipative coupling and gain-saturation non-linearity. \AN{The combination of these mechanisms, playing a crucial role in synchronization of driven-dissipative polariton condensate lattices, provides a multi-dimensional parameter space, allowing controllable switching between Kardar-Parisi-Zhang (KPZ) and Edward-Wilkinson (EW) dynamics.} Adiabatic elimination of fast density fluctuations reveals a wedge-shaped region in the complex hopping plane where the coarse-grained phase dynamics reduces to the Kardar-Parisi-Zhang (KPZ) equation. By computing high-resolution phase diagrams of the temporal and spatial scaling exponents we pinpoint the boundaries separating the KPZ domain from the Edwards-Wilkinson (EW) regime. Large-scale graphics processing unit (GPU) simulations of chains containing up to $N=2000$ condensates confirm these predictions: inside the wedge the exponents converge to $\beta_{N}=\textbf{0.329}(3)\!\approx\!1/3$ and $\chi_{N}=\textbf{0.504}(4)\!\approx\!1/2$, whereas outside it the dynamics moves away from KPZ and ultimately flows toward the EW fixed point, although finite system size and finite observation time may yield intermediate effective exponents.
\end{abstract}

\maketitle

\section{Introduction}\label{sec:intro}

Exciton-polaritons, composite bosonic quasiparticles formed by strong coupling between microcavity photons and quantum‑well excitons, constitute a versatile platform for exploring non‑equilibrium many‑body physics~\cite{Carusotto2013RMP}. One fascinating aspect is that polaritons can collectively achieve Bose-Einstein condensation~\cite{Kasprzak2006Nature}. Their exceptionally light mass and the finite photonic lifetime make polaritons intrinsically driven-dissipative: continuous non-resonant pumping feeds an incoherent reservoir that relaxes into a phase-coherent condensate while radiative decay converts the condensate field into a measurable optical signal~\cite{LittlewoodEdelman2017}.
\begin{figure}[t!]
    \centering
    \includegraphics[width=1\linewidth]{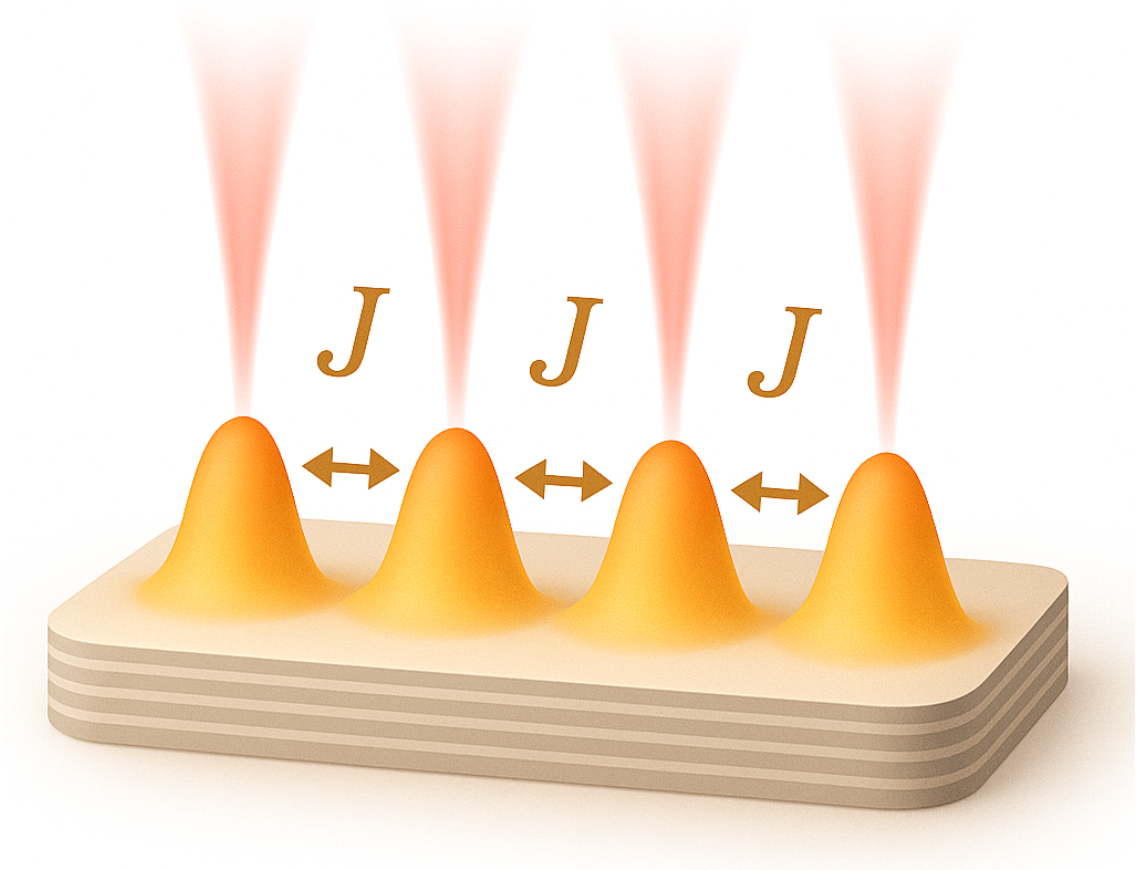}
\caption{Schematic illustration of a one-dimensional lattice of interacting polariton condensates created by an array of non-resonant \MM{optical} pump \MM{beams} \MM{(red)}. Each spot generates a localized condensate \MM{(orange)} whose tunnel coupling $J$ to its nearest neighbors has both coherent ($J_{\mathrm r}$) and dissipative ($J_{\mathrm i}$) components. These coupling strengths can be precisely tuned by adjusting the geometry and power of the pump spots.}
    \label{fig:setup}
\end{figure}
\AK{Such features raise the question of whether these polariton condensates can be used to emulate universality classes associated with non-equilibrium stochastic dynamics -- a key concept in critical phenomena whereby different systems exhibit the same scaling behavior near transitions and at large scales~\cite{stanley1971phase}. Two-dimensional arrays of exciton-polariton condensates have already been realized as fast XY analogue simulators, with phase locking across tens of sites on picosecond time scales~\cite{Berloff2017}. It also was shown that adding resonant forcing enables simulations of discrete Ising and $n$-state planar Potts models within the same platform~\cite{KalininBerloff}. Relatedly, Kibble-Zurek-type quenches in polariton lattices--where a control parameter is swept across a symmetry-breaking transition--offer access to universal defect-formation scaling. Complementary stochastic projected Gross-Pitaevskii-equation studies in ultracold atomic gases have elucidated noise-driven condensate growth and composite-defect formation after quenches~\cite{Liu2016PRA}.}

It turns out that the unique combination of tunable gain, losses, interactions, and intrinsic noise in polariton condensates hosts dynamical universality classes that have no analogue in systems at thermal equilibrium, where detailed balance precludes such non-linear scaling behaviour~\cite{Sieberer2016RPP}. A particularly important nonequilibrium fixed point is associated with the KPZ universality class originally introduced to describe classical surface growth~\cite{Kardar1986PRL}. Interested readers may consult Ref.~\cite{HalpinHealy1995PhysRep} for a thorough review of kinetic roughening and the KPZ universality class. In one dimension the KPZ equation  on $h(x,t)$, representing the height (or, more generally, the phase-like interface displacement) field, reads
\begin{equation}
  \partial_t h = \nu\partial_x^2 h + \frac{\lambda}{2}(\partial_x h)^2 + \mathcal{f}(x,t),
  \label{eq:KPZintro}
\end{equation}
where $\nu$ is an effective surface tension, $\lambda$ is the non-linear growth coefficient, and $\mathcal{f}(x,t)$ is a zero-mean Gaussian white noise. \MM{As for any stochastic growth model, the statistical properties of the surface $h(x,t)$ can be characterized by several power-law scalings with corresponding } universal exponents \(\beta=1/3\) (temporal growth \MM{of surface roughness}) and \(\chi=1/2\) (spatial roughness).  
Their exact values in $1\text{D}$ were first derived in Ref.~\cite{Kardar1986PRL}. In the next section, we define them by the scaling laws. Mapping the nonlinear Schr\"{o}dinger equation onto a noisy Burgers \MM{equation} allows using the condensates of one‑dimensional weakly interacting bosons for emulation of the KPZ dynamics~\cite{Kulkarni2013PRA}. A comprehensive account of these theoretical developments is also given in~\cite{Sieberer2016RPP}.  The exact results for generic Hamiltonian chains~\cite{vanBeijeren2012PRL} and driven spin models~\cite{Derrida1999JSP} confirmed the ubiquity of this scaling. \AK{Related KPZ scaling has also been identified in the finite-temperature hydrodynamics of isotropic Heisenberg spin chains. De Nardis \emph{et al.} formulated a two-mode nonlinear fluctuating hydrodynamics in which the local magnetization couples to an effective velocity field; this theory reproduces KPZ scaling of the dynamical spin structure factor while-consistent with time-reversal and inversion symmetries-yielding symmetric, quasi-Gaussian current fluctuations~\cite{DeNardis2023PRL}.} \AK{In another theoretical work~\cite{PhysRevX.5.011017} the authors have shown that in two-dimensional driven condensates the long-wavelength phase dynamics maps onto an (an)isotropic KPZ equation; in the isotropic case the KPZ nonlinearity destroys algebraic order at asymptotically long scales, whereas sufficiently strong anisotropy can restore an equilibrium-like Kosterlitz-Thouless phase.}

Early experimental confirmation beyond Eden-type growth came from measurements on turbulent liquid-crystal interfaces~\cite{Takeuchi2011}.
Direct experimental evidence of KPZ exponents in a single quasi‑1D polariton channel was reported by Fontaine \emph{et al.}~\cite{Fontaine2022Nature}. Temporal coherence deviations from linear theory, matching KPZ scaling, are shown in Ref.~\cite{Ji2015PRB}.  KPZ scaling was also investigated in optical parametric oscillator regime of microcavity polaritons by Ferrier \emph{et al.}~\cite{Ferrier}. \AK{More recently, Deligiannis \emph{et al.} showed that discretized two-dimensional polariton lattices can also exhibit KPZ roughening, even in the presence of topological defects, provided the lattice discretization modifies long-wavelength phase modes and defect dynamics~\cite{Deligiannis2022PRR,Deligiannis_2020}. 
This 2D scenario is complementary to the one we are going to study here: in one dimension there are no stationary spatial vortices; instead, phase slips manifest as space-time vortex-antivortex events~\cite{He2017PRL}, which can drive a long-time crossover from KPZ stretched-exponential temporal coherence to a simple exponential decay beyond an exponentially large scale.}
Scaling properties of one-dimensional driven-dissipative condensates were also numerically studied in Ref.~\cite{He2015PRB}. These results raise a fundamental and a practical question: under which conditions does a lattice of multiple interacting condensates support KPZ scaling in its phase field? Besides its conceptual interest, answering this question is crucial for analogue simulation schemes that rely on coherent networks of polariton condensates. In practice, the KPZ or EW character of the phase dynamics sets the attainable coherence length and computation time of a polariton-based analogue simulator; our phase diagrams provide a concrete recipe for choosing pump geometry and inter-site coupling so that the device operates either in a highly coherent (EW) regime or, when desired, in a noise-
dominated (KPZ) regime for stochastic sampling tasks.

Moroney and Eastham \cite{Moroney2023PRB, Moroney2021} considered a one‑dimensional chain of exciton-polariton condensates with complex nearest‑neighbour coupling $J\equiv J_{\mathrm r}+iJ_{\mathrm i}$ and showed that the imaginary component, which stems from reservoir‑mediated dissipative tunnelling, can destabilise the synchronous phase via the proliferation of space-time vortices~\cite{Moroney2023PRB}.  Their analytical bound suggests that KPZ behaviour should survive provided that $|J_{\mathrm i}|$ is sufficiently small compared with $J_{\mathrm r}$.  
\AN{At the same time, the dissipative reservoir-mediated coupling mechanism is dominant in establishing the synchronous coherent polariton condensate lattice state and can be optically controlled as well as the conservative coupling \cite{Alyatkin2020,Alyatkin2021}.
This requires a systematic exploration of the stochastic dynamics regimes on the complex plane of $(J_{\mathrm r},J_{\mathrm i})$, which is yet missing.}

In the present work we derive the effective phase dynamics of an 
optically induced polariton lattice starting from a microscopic driven-dissipative Gross-Pitaevskii equation (GPE) that includes stochastic reservoir kinetics. In Fig.~\ref{fig:setup}, we schematically present the model system. It represents a one-dimensional array of exciton-polariton condensates created by a non-resonant optical pumping. The condensates lose polaritons due to their radiative decay. They are replenished due to the stimulated scattering of excitons from incoherent reservoirs. We account for the coherent and dissipative couplings of the condensates~\cite{Aleiner2012PRB, LagoudakisPRX}. By adiabatically eliminating density fluctuations we obtain a stochastic nonlinear equation for the lattice phase which reduces to the KPZ form within a finite wedge of coupling parameters.  We corroborate this analytical criterion by extensive numerical simulations of chains with up to $2000$ condensates -- matching present experimental capabilities -- and analyse finite‑size effects and the crossover to Edwards-Wilkinson scaling. 

The remainder of the paper is organised as follows.  Section~\ref{sec:model} presents the driven-dissipative GPE with reservoir, its reduction to a discrete lattice model and the derivation of the KPZ coefficients. There, we also detail the numerical methods used in our simulations.  Section~\ref{sec:results} discusses the resulting phase diagrams and scaling exponents.  Section~\ref{sec:concl} summarises our findings and outlines perspectives of extension of the model for description of KPZ-simulators based on higher‑dimensional lattices of polariton condensates.

\section{Microscopic model and KPZ reduction}\label{sec:model}

In this section we formulate the driven-dissipative Gross-Pitaevskii description appropriate for a one‑dimensional array of optically trapped exciton-polariton condensates, derive its discrete lattice representation, and outline the continuum limit leading to Kardar-Parisi-Zhang (KPZ) scaling.

\subsection{Driven-dissipative Gross-Pitaevskii equation}\label{subsec:ddgpe}
We consider a high-quality planar semiconductor microcavity pumped non-resonantly by a mask-shaped laser profile that generates a set of $N$ spatially separated exciton reservoirs.  The coherent polariton order parameter $\Psi(\mathbf r,t)$ interacts with the incoherent reservoir density $n_R(\mathbf r,t)$ and obeys the driven-dissipative Gross-Pitaevskii equation (ODGPE)~\cite{Wouters2007PRL}
\begin{subequations}\label{eq:ODGPE}
\begin{align}
  i\hbar\,\partial_t \Psi &=
    \Bigl[-\frac{\hbar^2\nabla^2}{2m}+\frac{i\hbar}{2} Rn_R - \frac{i\hbar}{2}\Gamma_0\abs{\Psi}^2 - \frac{i\hbar}{2}\gamma_C \nonumber \\ &\qquad\qquad\qquad\qquad+ \hbar\alpha|\Psi|^2\Bigr]\Psi+\hbar\xi,
    \label{eq:psi_cont}\\
  \partial_t n_R &=
    P(\mathbf r)-\bigl(\gamma_R+R|\Psi|^2\bigr)n_R.
    \label{eq:res_cont}
\end{align}
\end{subequations}
Here $m$ is the lower-polariton effective mass, $\alpha$ is the polariton-polariton interaction constant, $\gamma_C$ ($\gamma_R$) is the decay rate of the condensate (reservoir), $R$ is the stimulated-scattering rate from the reservoir into the condensate, \DN{$\Gamma_0$ is the gain saturation,} and $P(\mathbf r)$ denotes the spatially patterned non-resonant pump.  Fluctuations are represented by complex Gaussian white noise $\xi$ with correlator $\langle\xi^{*}(\mathbf r,t)\xi(\mathbf r',t')\rangle = 2D_0\,\delta(\mathbf r-\mathbf r')\delta(t-t')$.

When the reservoir \MM{dissipates} much more slowly than the condensate, $\gamma_R\!\ll\!\gamma_C$, the characteristic reservoir timescale $\tau_R=\gamma_R^{-1}$ greatly exceeds the condensate timescale $\tau_C=\gamma_C^{-1}$.  In this adiabatic limit $n_R$ can be treated as quasi-static: $n_R(\mathbf r,t)\simeq n_R^{(0)}(\mathbf r)$.  Dropping Eq.~\eqref{eq:res_cont} and replacing the gain term \DN{$\tfrac{i\hbar}{2} \qty[R n_R-\gamma_C]$} in Eq.~\eqref{eq:psi_cont} by an effective pump  \DN{$P_C(\mathbf r)= R n_R^{(0)}-\gamma_C$}, one obtains the reduced form \DN{(similar to \cite{Moroney2021})}
\begin{align}
  i\hbar\,\partial_t \Psi &=
    \Bigl[-\frac{\hbar^2\nabla^2}{2m}+\frac{i\hbar}{2} P_C -\frac{i\hbar}{2}\Gamma_0\abs{\Psi}^2 \nonumber \\ &\qquad\qquad\qquad\qquad\qquad\qquad+ \alpha|\Psi|^2\Bigr]\Psi+\xi,
    \label{eq:psi_cont1}
\end{align}
This simplified equation is used in the subsequent derivation of the discrete lattice model.

\subsection{Discrete lattice model}\label{subsec:lattice}

The patterned pump is arranged as an array of $N$ identical Gaussian spots centred at positions $x_n = n a$ ($n = 1,\dots, N$) with lattice period $a$.  Each spot supports a normalised single-particle orbital $\phi_n(\mathbf r)$ whose spatial overlap is appreciable only with its nearest neighbours.  Representing the condensate field as
\begin{align}
  \Psi(\mathbf r,t)=\sum_{n=1}^{N}\psi_n(t)\,\phi_n(\mathbf r),
  \label{eq:expansion}
\end{align}
and projecting the reduced ODGPE~\eqref{eq:psi_cont1} onto $\phi_n$ yields a discrete driven-dissipative Gross-Pitaevskii equation for the complex amplitudes $\psi_n(t)$.

After retaining nearest-neighbour hopping and adopting the notation introduced earlier, the evolution equation that is actually integrated reads~\cite{Moroney2023PRB}
\begin{align}
  i\hbar\dot\psi_n &=
    \Bigl[E_0+\frac{i\hbar}{2}P-\frac{i\hbar}{2}\gamma_0\abs{\psi_n}^2+\alpha\abs{\psi_n}^{2}\Bigr]\psi_n\nonumber\\
    &\qquad\qquad\qquad\qquad\qquad+\sum_{k=\langle n\rangle}J\,\psi_k
    +\hbar\xi_n ,
    \label{eq:psi_disc}
\end{align}
where $E_0$ is the on-site kinetic energy, $P$ the effective gain supplied by the pump, \DN{$\gamma_0$ the gain saturation,} $\alpha$ the on-site interaction strength, and $J=J_{\mathrm r}+iJ_{\mathrm i}$ the complex nearest-neighbour coupling introduced in Sec.~\ref{subsec:KPZ_connection}.  The stochastic term $\xi_n(t)$ is the lattice counterpart of the white noise $\xi(\mathbf r,t)$ defined in Sec.~\ref{subsec:ddgpe}.  Equation~\eqref{eq:psi_disc} is solved numerically for chains containing up to $N = 2000$ condensates.

\subsection{Connection to KPZ universality}\label{subsec:KPZ_connection}

All numerical data reported below are obtained by integrating the
complex-field lattice equation~\eqref{eq:psi_disc}.
To interpret those results it is instructive to figure out how the same
microscopic model reduces, at long wavelengths, to the
one-dimensional Kardar-Parisi-Zhang (KPZ) equation that governs the
universal fluctuations of the condensate phase.

We decompose the lattice field into amplitude and phase
\begin{align}
  \psi_i(t)=\sqrt{n_i(t)}\,e^{i\theta_i(t)},
  \label{eq:polar}
\end{align}
where $n_i$ and $\theta_i$ are the on-site density and phase.
Density fluctuations relax on the time-scale of
$\tau_n\!\sim\!(\gamma_0)^{-1}$, which is short compared to the
time-scale of phase diffusion.
Eliminating $\delta n_i=n_i-n$ adiabatically and expanding finite
differences to the second order in the lattice spacing
$a$ one arrives, in the continuum limit $x=i a$, at
\begin{align}
  \partial_t\theta(x,t)=
      \nu\,\partial_x^2\theta
    +\frac{\lambda}{2}\bigl(\partial_x\theta\bigr)^2
    +\zeta(x,t),
  \label{eq:KPZ}
\end{align}
the standard one-dimensional KPZ equation.
The effective diffusion and non-linear coefficients read
\begin{align}
  \nu=\frac{2 a^{2}\!\bigl(J_{\mathrm i}-\eta J_{\mathrm r}\bigr)}{\hbar},
  \qquad
  \lambda=-\frac{a^{2}\!\bigl(J_{\mathrm r}+\eta J_{\mathrm i}\bigr)}{\hbar},
  \label{eq:Dlambda}
\end{align}
\AK{with $\eta=2\alpha/(\hbar\gamma_0)$ quantifies the interaction-to-amplitude-relaxation ratio (with $-\tfrac{i\hbar}{2}\gamma_0|\psi|^2$ setting the density relaxation rate) and
$J=J_{\mathrm r}+iJ_{\mathrm i}$.
The term $\zeta(x,t)$ is Gaussian white noise, inherited from
$\xi_i(t)$.}

Equation~\eqref{eq:KPZ} belongs to the strong-coupling KPZ
universality class provided the diffusion constant is positive,
\begin{align}
  J_{\mathrm i}>\eta\,J_{\mathrm r}.
  \label{eq:wedge}
\end{align}
When, in addition, the non-linear coefficient $\lambda$ vanishes
($J_{\mathrm r}+\eta J_{\mathrm i}=0$) the dynamics crosses over to the
Edwards-Wilkinson universality class with critical exponents
$\beta=1/4$ and $\chi=1/2$.
Although this coarse-grained reduction is not used in the simulations,
condition~\eqref{eq:wedge} provides a practical guideline for selecting
hopping parameters that display KPZ scaling in the full complex-field
model. \AK{Let us stress here that the adiabatic elimination presented above hinges on the time-scale separation $\gamma_0 n_0 \gg \nu k^2,|\lambda|k^2$, rather than on weak $\alpha$. All exponents reported below are obtained from the full complex-field model~\eqref{eq:psi_disc}. The KPZ reduction is used only for analytic guidance (the wedge $J_{\mathrm i}>\eta J_{\mathrm r}$). Larger $\eta$ rotates this wedge without changing the universality.}

\subsection{Numerical integration protocol}\label{subsec:numerics}

To capture nonlinear effects beyond a phase-only description we integrate the full complex-field equation~\eqref{eq:psi_disc}. 
In our simulations, the chain length is varied up to $N=2000$, enabling a systematic finite-size study. 

The stochastic differential equations are treated with a semi-implicit Euler-Maruyama algorithm. Space-time first-order coherence is evaluated on the lattice
\begin{align}
  g^{(1)}(\Delta x,\Delta t)\!\!=\!\!
  \frac{\bigl\langle\psi^{*}(x_0,t_0)\,\psi(x',t')\bigr\rangle}
       {\expval{\abs{\psi(x_0, t_0)}\abs{\psi(x', t')}}},
  \label{eq:g1}
\end{align}
where $x'=x_0+\Delta x$, $t'=t_0+\Delta t$, and $\langle\cdots\rangle$ denotes an ensemble average. The coordinate-time pair $(x_0,t_0)$ is chosen deeply inside the chain after transients have decayed.  

To improve the ensemble statistics, we calculate the ensemble average over $M$ different systems of $N$ condensates  by considering chain of condensates with length $N\cdot M$, which consists of $M$ independent sub-chains of the length $N$ each. Time evolution and average values of this ensemble were calculated on a GPU using Compute Unified Device Architecture (CUDA) library.

The critical exponents are then extracted from the following scaling relations
\begin{align}
  C(\Delta x = 0, \Delta t) \equiv -2\ln\bigl|g^{(1)}(0,\Delta t)\bigr|&\propto\Delta t^{2\beta},\\
  C(\Delta x, \Delta t = 0) \equiv -2\ln\bigl|g^{(1)}(\Delta x,0)\bigr|&\propto\Delta x^{2\chi}, \label{eq:g1_x}
\end{align}
with the exponents $\beta$ and $\chi$ obtained by least-squares fits over two decades in $\Delta t$ or $\Delta x$ on log-log axes.

Throughout we set $\hbar = 1$ and measure energy in units of \DN{the gain saturation rate} $\gamma_0$; physical time is therefore rescaled as $\tau = t\gamma_0$. A dimensionless time step of $\Delta\tau = 0.1$ was adopted, and convergence was explicitly verified by repeating the simulations with smaller steps. The linear\MM{ly polarized} pump is fixed at $P/\gamma_0 = 1$, and the interaction-to-damping ratio is kept weak, $\eta = 2\alpha/(\hbar\gamma_0) = 10^{-4}$, as appropriate for large-area microcavities. Also note, that the on-site kinetic energy $E_0$ does not affect first-order correlator $g^{(1)}$ and only adds a constant drift in a phase temporal evolution $(\theta_i(t)\sim -E_0t/\hbar)$, and is therefore set to zero without loss of generality. Within this parameter set we scan the complex coupling amplitude $J$ in both real and complex directions. This variation can be achieved by changing the pump geometry. Experimental engineering of spatial coherence and couplings in polariton lattices appears in Ref.~\cite{Topfer2021Optica}.

The noise strength $D$ can be tuned to change the coherence time of the system but it should be low enough for condensate fluctuation being small ($\Delta n_i \ll n_i$). This condition is required by the adiabatic elimination that leads to the KPZ equation~\eqref{eq:KPZ}. We choose $D$ so that the root-mean-square density fluctuation equals 5 \% of the steady-state density obtained in the noiseless limit
\begin{figure}[b]
  \includegraphics[width=\columnwidth]{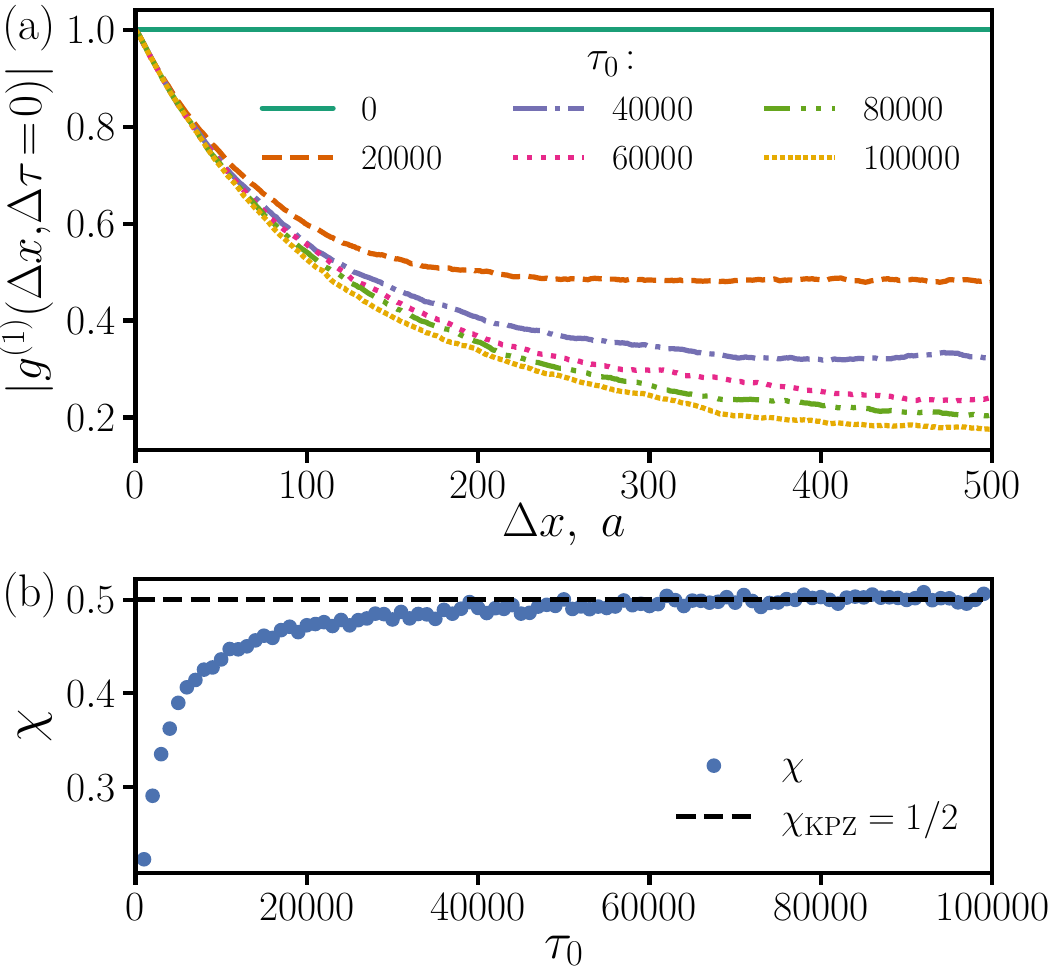}
\caption{Relaxation of the spatial coherence toward the KPZ regime. Panel (a): equal-time correlator $g^{(1)}(\Delta x,0)$ computed for different waiting times $\tau_0=t_0\gamma_0$:  $0$, $2\times 10^4$,  $4\times10^4$, $6\times 10^4$, $8\times10^4$, and $10^5$. Panel (b): roughness exponents $\chi$ extracted from the large-$\Delta x$ tails. For $\tau_0 = 5\times 10^{4}$ the correlator has completely lost memory of the uniform initial state and the exponent has approximately converged to the KPZ value $\chi \simeq 1/2$ (horizontal black dashed line), justifying the choice of this waiting time for all subsequent analysis. The selection of $\tau_0$ is crucial to ensure that transient effects from the initial coherent phase are fully relaxed, so that subsequent measurements reflect the true KPZ scaling regime.}
\label{fig:cor_beh}
\end{figure}
\begin{figure}[t!]
\includegraphics[width=\columnwidth]{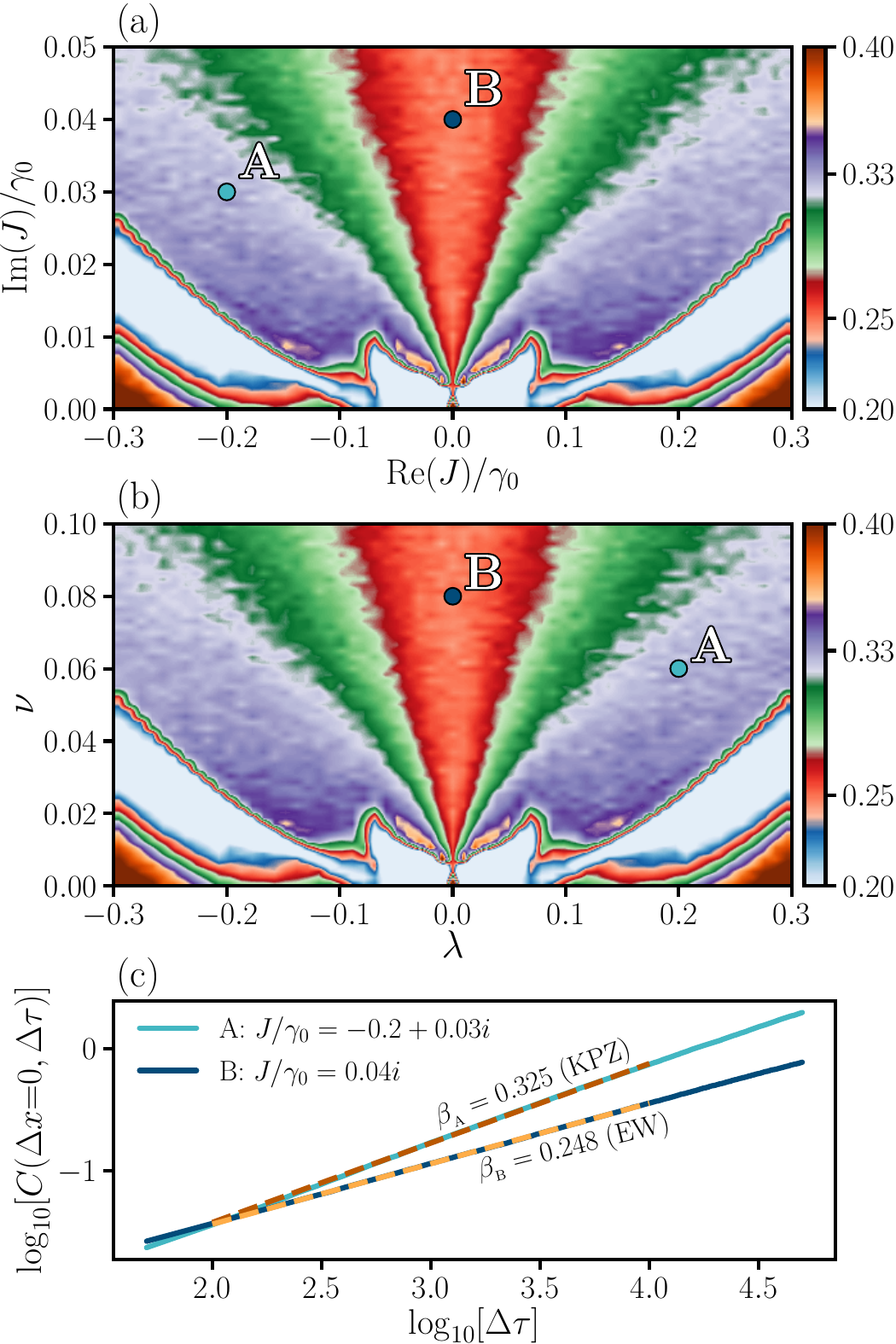}
\caption{%
(a)~Colour map of the temporal growth exponent $\beta$ extracted from $-2\ln|g^{(1)}(0,\Delta t)|\propto\Delta t^{2\beta}$ in the plane of real and imaginary parts of the hopping amplitude $(J_{\mathrm r},J_{\mathrm i})$. The data correspond to interaction $\eta=10^{-4}$, chain length $N=500$, and ensemble size $M=10\,000$. (b)~The same data represented in the variables $\lambda=-(J_{\mathrm r}+\eta J_{\mathrm i})$ and $\nu = 2(J_{\mathrm i}-\eta J_{\mathrm r})$ that enter the coarse-grained KPZ equation (recall that $\hbar=1$ and $a=1$). Because the working value $\eta\!=\!10^{-4}\ll1$, the mapping produces, to excellent accuracy, a mirror reflection of panel (a) with respect to the vertical axis; we include it to make the KPZ phase boundary immediately visible in the conventional $(\lambda,\nu)$ language. (c)~Log-log plots of $-2\ln|g^{(1)}(0,\Delta t)|$ for the two marked points A (light blue pint on panel (a) and (b)) ($J/\gamma_0=-0.2+0.03i$), where KPZ scaling is well pronounced, and B (deep blue point on both upper panels) ($J/\gamma_0=0.04i$), for which EW scaling is expected.  Linear fits (dashed) yield $\beta_{_\text{A}}\simeq0.325\;(1/3)$ for A, characteristic of the KPZ class, and $\beta_{_\text{B}}\simeq0.248\;(1/4)$ for B, consistent with Edwards-Wilkinson (EW) scaling.  Panel (c) was obtained for a larger ensemble ($M=100\,000$) to suppress statistical noise.}
\label{fig:phase}
\end{figure}
\begin{align}
  \sqrt{D} = 0.05\,\frac{P + 4J_{\mathrm i}}{\sqrt{2\gamma_0\Delta\tau}} .
\end{align}
An initial condition for $\psi$ does not affect steady state solution at $t\to\infty$ but it defines how fast the system's dynamics enters the specific universality class regime. We are using equal-phase initial conditions $\psi_i(0)=\psi_j(0)=const$. With such initial conditions the system is spatially correlated at $t=0$:
$g^{(1)}(\Delta x, 0)\big|_{t_0=0} = 1.$ So KPZ universality class characterized by spatial correlations of Eq.\eqref{eq:g1_x} can be established only after some time of phase fluctuations.  All statistical sampling begins after a waiting time of $\tau_0 = 5\times10^{4}$, which is well beyond this transient.

\section{Results}\label{sec:results}

In this section, we explore how the critical exponents extracted from Eq.~\eqref{eq:psi_disc} depend on the complex hopping amplitude $J=J_{\mathrm r}+iJ_{\mathrm i}$, on the interaction-damping ratio $\eta=2\alpha/(\hbar\gamma_0)$, and on the chain length $N$. Unless stated otherwise the interaction is fixed at the experimentally relevant value $\eta = 10^{-4}$ and the noise strength is chosen as
described in Sec.~\ref{subsec:numerics}.

First, as argued at the end of Sec.~\ref{subsec:numerics}, the KPZ scaling regime is reached only after an initial transient. Before running the large-scale simulations we therefore determine a waiting time $t_0$ beyond which correlations can be analysed safely. Fig.~\ref{fig:cor_beh} shows the equal-time spatial correlator $g^{(1)}(\Delta x,0)$ and the roughness exponent $\chi$ extracted from it for several choices of $t_0$. The data demonstrate that the value adopted in the following, $\tau_0 = t_0\gamma_0 = 5\times10^{4}$, is already sufficient for the influence of the uniform initial condition to vanish. Therefore, this time moment is used as the starting point for all statistical analyses.

\begin{figure}[t!]
  \includegraphics[width=1\columnwidth]{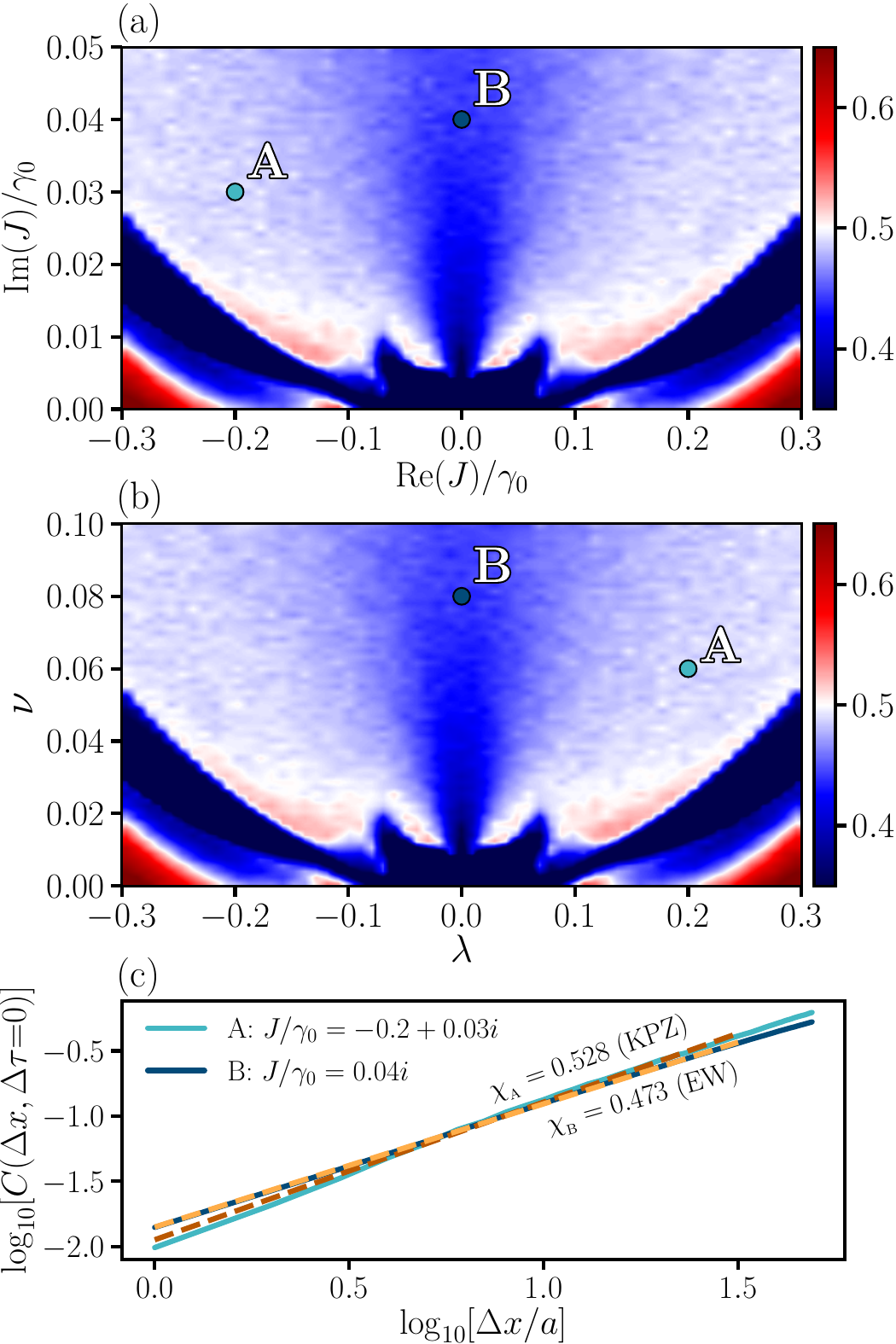}
  \caption{Spatial roughness exponent $\chi$. (a)~Colour map of $\chi$ in the $(J_{\mathrm r},J_{\mathrm i})$ plane for interaction $\eta=10^{-4}$, chain length $N=500$, and $M=10\,000$ noise realizations. (b)~Same data expressed in the KPZ variables $\lambda$ and $\nu$; because $\eta\ll1$ the image is essentially a mirror reflection of panel~(a). (c)~Log-log plot of $-2\ln|g^{(1)}(\Delta x,0)|$ for points A ($J/\gamma_0=-0.2+0.03i$) and B ($J/\gamma_0=0.04i$), where a linear fit (dashed lines) gives $\chi_{_\text{A}}\simeq0.528\approx1/2$ and $\chi_{_\text{B}}\simeq0.473\approx1/2$, in agreement with KPZ and EW scaling.}
  \label{fig:chi}
\end{figure}
\begin{figure}[t!]
  \includegraphics[width=\columnwidth]{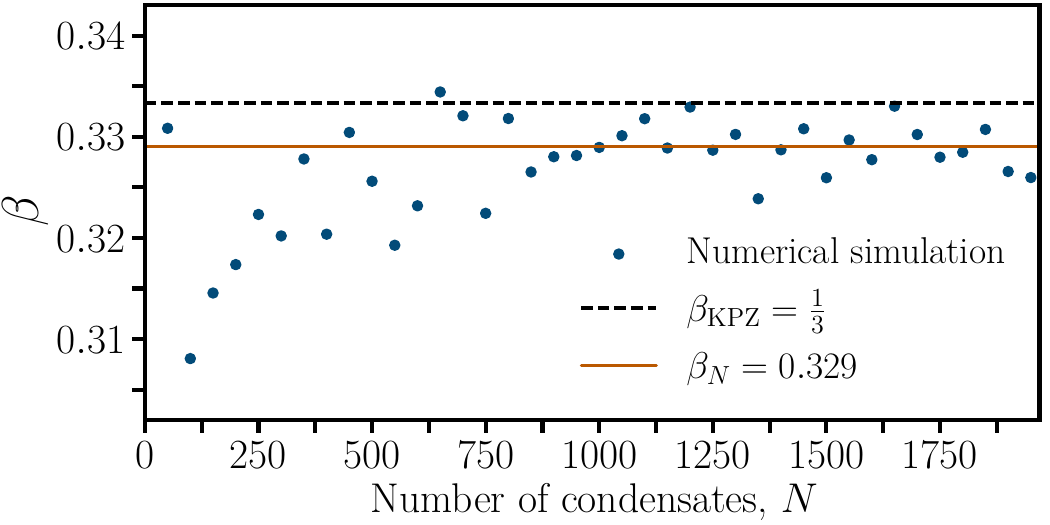}\\
  \includegraphics[width=\columnwidth]{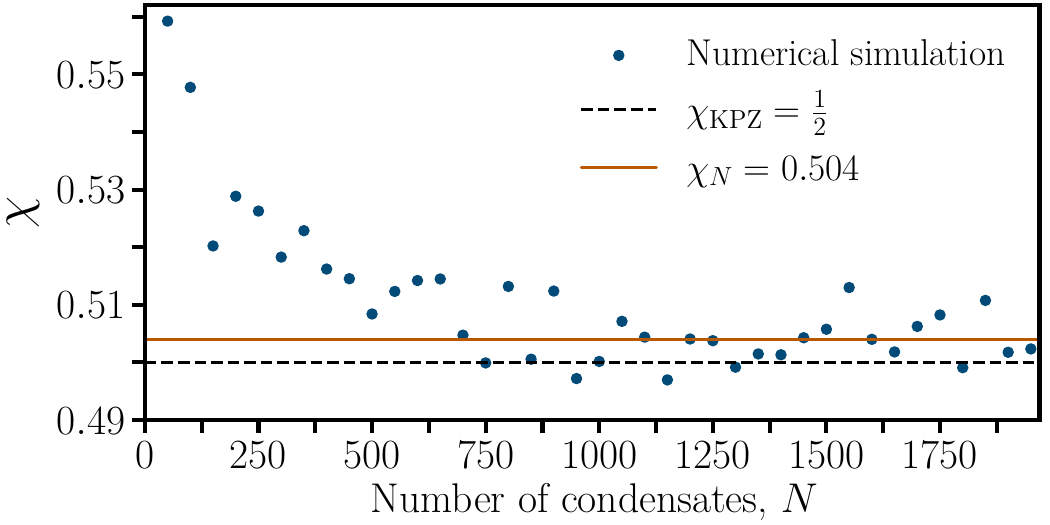}
\caption{Dependencies of the critical exponents on the size of the system at $J/\gamma_0=-0.2+0.03i$ (point~\textsf{A} in Fig.~\ref{fig:phase} and Fig.~\ref{fig:chi}). Top panel: temporal growth exponent $\beta$ as a function of the chain length~$N$; the horizontal dashed line marks the KPZ value for exponent $\beta_{\mathrm{KPZ}}=1/3$, while the solid orange line shows the numerical plateau $\beta_N = 0.329\pm0.003$ obtained by averaging the last twenty points ($N\!\ge\!1000$). Bottom panel: spatial roughness exponent $\chi$ versus~$N$. The dashed line indicates the KPZ value $\chi_{\mathrm{KPZ}}=1/2$ and the solid orange line shows the numerical estimate $\chi_N = 0.504\pm0.004$.   Each point represents an average over $M=10^{4}$ independent noise
realizations.}
  \label{fig:chi_and_beta}
\end{figure}

Now let us emphasize the main result of our work. Fig.~\ref{fig:phase}(a) shows the colour map of the temporal growth exponent $\beta$ in the complex $J$ plane. For the chosen weak interaction the KPZ criterion $J_{\mathrm i} > \eta J_{\mathrm r}$ (see sec.~\ref{subsec:KPZ_connection}) nearly corresponds to the upper half-plane $J_{\mathrm i}>0$, except for the vertical axis $J_{\mathrm r}=0$, where the non-linear coefficient $\lambda$ vanishes and Edwards-Wilkinson (EW) scaling should be observed. This analytical expectation is confirmed by the simulations: a \MM{red region} with $\beta\!\simeq\!1/4$ (EW class) appears along $J_{\mathrm r}=0$ axis, while the bulk of the upper half-plane is violet with $\beta\!\simeq\!1/3$ (KPZ class). The green area represents a crossover region: the fitted exponents there differ from both KPZ and EW because the required crossover scales exceed our simulation window.  In the infinite-size, long-time limit these points are expected to flow to the EW fixed point.
Close to the real axis the fit of $-2\ln|g^{(1)}(0,\Delta t)|\propto\Delta t^{2\beta}$ becomes unreliable (light blue region) because the diffusive coefficient $\nu$ is very small, and the statistical spread of the phase grows, requiring much more noise realizations for convergence. Fig.~\ref{fig:phase}(b) replots exactly the same data in $(\lambda,\nu)$ variables of the KPZ equation using the relations $\lambda=-(J_{\mathrm r}+\eta J_{\mathrm i})$ and $\nu = 2(J_{\mathrm i}-\eta J_{\mathrm r})$, where we allow $a=1$. With our experimentally realistic value $\eta=10^{-4}$ the transformation reduces essentially to a mirror reflection with respect to the vertical axis, so panel (b) is the left-right image of panel (a); nevertheless it makes evident the location of the KPZ wedge in the $(\lambda,\nu)$ plane. Fig.~\ref{fig:phase}(c) illustrates the two universality classes explicitly. For the point $J/\gamma_0=-0.2+0.03i$ the $\log-\log$ plot of $-2\ln|g^{(1)}(0,\Delta t)|$ yields a slope $2\beta = 2/3$, characteristic of KPZ scaling, whereas for the purely imaginary coupling $J/\gamma_0 = 0.04i$ the slope is $2\beta = 1/2$, as predicted by EW theory.

Fig.~\ref{fig:chi} shows the counterpart of Fig.~\ref{fig:phase}(a) for the spatial roughness exponent $\chi$, extracted from the scaling law
$-2\ln|g^{(1)}(\Delta x,0)|\propto\Delta x^{2\chi}$. Panel~(a) shows the colour map in the original variables $(J_{\mathrm r},J_{\mathrm i})$. Over nearly the entire upper half-plane ($J_{\mathrm i}>0$) the simulations predict $\chi\simeq 0.5$, in the excellent agreement with the one-dimensional KPZ value $\chi_{\rm KPZ}=1/2$. Close to the real axis, however, a narrow red band appears where $\chi$ cannot be determined reliably; this is the same region where Fig.~\ref{fig:phase}(a) showed large uncertainties in the exponential factor~$\beta$. In the limit $J_{\mathrm i}\to 0$ the effective diffusion constant $\nu$ vanishes, phase gradients become extremely rough, and the number of noise realizations required for converged statistics grows faster than our ensemble size. The light blue EW region at $J_{\mathrm r}=0$ in Fig.~\ref{fig:chi} likewise originates from the finite system size. We note that for the spatial correlator the roughness factor $\chi$ takes the same universal value in both the KPZ and EW classes. When the real part of the coupling is set to zero the numerical convergence deteriorates and finite--size effects of the simulated chain become much more pronounced. For completeness we re-plot exactly the same data in panel~(b) using the KPZ parameters $\lambda=-(J_{\mathrm r}+\eta J_{\mathrm i})$ and $\nu = 2(J_{\mathrm i}-\eta J_{\mathrm r})$. Because the working value $\eta=10^{-4}$ is extremely small, it amounts practically to a mirror reflection of panel~(a) about the vertical axis. This alternative representation makes the location of the KPZ wedge in the conventional $(\lambda,\nu)$ plane quite clear. Panel~(c) illustrates the scaling at the representative point A (marked in panels~a,b). The log-log plot of $-2\ln|g^{(1)}(\Delta x,0)|$ versus $\Delta x/a$ shows a slope $2\chi\simeq0.966$, i.e. $\chi\simeq0.483$, that is close to the KPZ value given the numerical accuracy imposed by the size of the finite system.

Fig.~\ref{fig:chi_and_beta} (top panel) tracks the growth exponent $\beta$ as a function of chain length for the representative coupling $J/\gamma_0=-0.2+0.03i$. Starting from $\beta\approx 0.31$ at $N=100$, the value rises steadily and saturates at the KPZ benchmark $\beta_{\rm KPZ}=1/3$ once $N\gtrsim 1000$. The bottom panel of Fig.~\ref{fig:chi_and_beta} shows the companion finite-size flow of $\chi$, which approaches $1/2$ on the same length scale. These trends \AK{suggest} that the EW-like window shrinks as $N^{-1}$ and collapses onto the line $J_{\mathrm r}=0$ in the thermodynamic limit, as it is expected from the analytical criterion $J_{\mathrm i}>\eta J_{\mathrm r}$.

We note, however, that a residual spread of individual data points is still visible in both panels. This spread lies within the numerical accuracy of the simulation and reflects (i)~the finite number of noise realizations, (ii)~the stochastic nature of the least-squares fits, and (iii)~the intrinsic difficulty of extracting scaling exponents from a highly nonlinear driven-dissipative system.

To provide definite numerical benchmarks for the most favorable parameter region highlighted in Figs.~\ref{fig:phase}~(c) and
\ref{fig:chi}~(c), we quote final estimates obtained as the average over the last twenty data points in Fig.~\ref{fig:chi_and_beta}, i.e. over chains with $N\!\ge\!1000$ where additional condensates do not improve the statistics further.
The resulting values
\begin{align}
  \beta_{N, \rm KPZ} = 0.329(3), 
  \qquad
  \chi_{N, \rm KPZ}  = 0.504(4)
\end{align}
are drawn as solid orange guide lines in the figure and are taken as the final numerical KPZ estimates of our model.

In summary, both critical exponents approach their KPZ benchmarks once the chain hosts only a few hundred condensates -- an array size that can be realized with current spatial-light-modulator techniques and state-of-the-art microcavities. The convergence is monotonic: as $N$ increases the growth exponent $\beta$ rises from its finite-size value and locks onto $1/3$, while the roughness exponent $\chi$ descends toward $1/2$ on the same length scale, indicating that phase gradients have reached their asymptotic strong-coupling statistics. The modest residual offsets observed very close to $J_{\mathrm i}=0$ should not be interpreted as a failure of KPZ scaling. They stem from the vanishing diffusion constant $\nu\propto J_{\mathrm i}$, which slows down the equilibration of long-wavelength modes and amplifies statistical noise.
Longer simulation times or a larger ensemble of noise realizations would reduce this bias further, but the overall trends remain unchanged.

\section{Conclusion}\label{sec:concl}

We have combined analytical and numerical approaches to chart the conditions under which one-dimensional arrays of driven-dissipative exciton-polariton condensates emulate the Kardar-Parisi-Zhang universality class equations. Starting from a stochastic Gross-Pitaevskii description that retains both gain saturation and reservoir noise, we show that adiabatic elimination of density fluctuations yields an effective KPZ equation for the lattice phase whenever the inequality $J_{\mathrm i}>\eta J_{\mathrm r}$ is satisfied. Extensive GPU simulations -- covering chain lengths up to two thousand sites and ensembles of $10^{4}$ noise realisations -- corroborate this criterion: the growth and roughness exponents approach the KPZ benchmarks, $\beta=1/3$ and $\chi=1/2$, once the system contains a few hundred condensates. For the coupling $J/\gamma_0=-0.2+0.03 i$ we obtain final finite-size estimates of $\beta_{N}=0.329(3)$ and $\chi_{N}=0.504(4)$, providing the specific numerical targets for experiments. Our findings reconcile the KPZ scaling observed in single polariton wires~\cite{Fontaine2022Nature} with the space-time-vortex scenario proposed for coupled condensates~\cite{Moroney2023PRB}, and delineate the parameter window where coherent polariton networks can serve as analogue simulators of non-equilibrium surface growth. Further studies are required to the present analysis to two-dimensional lattices, explore the impact of disorder and detuning, and investigate whether engineered pump geometries can stabilise higher-dimensional KPZ fixed points in driven-dissipative quantum fluids.
\AN{The discovered mechanism of control over the stochastic dynamics regime present a particular interest for optimizing characteristic synchronization and coherence times in polariton condensate analogue simulators.}

\textit{Acknowledgments}.-- The research is supported by the Russian Science Foundation under Grant No. 25-12-00135 (analytical derivation) and by Ministry of Science and Higher Education of the Russian Federation (Goszadaniye) Project No. FSMG-2023-0011 (numerical simulation). The work of A.K. is supported by the Icelandic Research Fund (Ranns\'oknasj\'o{\dh}ur, Grant No.~2410550).

\bibliography{kpz_refs}

\end{document}